# Enhancing Evolutionary Solver Efficiency for NP-Hard Single Machine Scheduling Problems

**Mohammed Al-Romema[1], Mohammed A. Makarem[2]**

[1]Department of Industrial & Systems Engineering, King Fahd University of Petroleum and Minerals, Dhahran, Saudi Arabia
Email: *mohammadalromaima1992[at]hotmail.com*

[2]Department of Computer Engineering, King Fahd University of Petroleum and Minerals, Dhahran, Saudi Arabia
Email: *Makarem.a.Mohammed[at]gmail.com*

**Abstract:** *The study explores the optimization of evolutionary solver parameters for minimizing total tardiness in single machine scheduling, an NP-hard problem with zero ready times included. It investigates various parameter combinations, including population sizes, mutation rates, and a constant convergence rate, both above and below default values. The aim is to enhance the solver's effectiveness in addressing this complex challenge. The findings contribute to improving scheduling efficiency in manufacturing and operations management contexts.*

## 1. Introduction

In the contemporary manufacturing landscape, characterized by constrained resources, escalating consumer demands, and intense competition both domestically and globally, the role of scheduling emerges as a pivotal function within any manufacturing enterprise. In the context of an increasingly globalized and international business environment, the imperatives of cost reduction and profit maximization continue to drive the strategic decisions of manufacturing companies. The impact of manufacturing scheduling on a company's operational performance, and consequently its sustainability, is significant, particularly when considering the correlation between tardiness and operational expenses[1]. This study aims to delve into the optimization of parameters for an evolutionary solver tasked with addressing the challenge of minimizing total tardiness in single machine scheduling, a problem denoted as $1||\Sigma T_j$ when using Excel solver. It is assumed that all jobs have zero arrival time. The objective is to strategically sequence jobs to minimize total tardiness ($\Sigma T_j$), which is determined by the formula $T_j = \text{Max}\{C_j - d_j, 0\}$, where $T_j$ is the tardiness of job $j$, $C_j$ denotes the completion time, and $d_j$ represents the due date. This research specifically focuses on modifying certain key parameters - population size, mutation rate, and convergence - setting them both above and below the default values. This approach is undertaken to establish a robust and general set of parameters that effectively contribute to the minimization of total tardiness, addressing the core objective function of this study.

## 2. Problem Statement

The article aims to optimize the parameters of an evolutionary solver to efficiently address the NP-hard problem of minimizing total tardiness in single-machine scheduling scenarios. It's study posits a scenario where a series of jobs, labeled as $j = 1, 2,..., n$, are available for processing on a single machine. Each job $j$ has an associated processing time, denoted as $p_j$, and a due date, represented as $d_j$. Once initiated, a job must be completed on the machine without interruption. It is important to note that the machine is designed to handle only one operation at a time. The concept of tardiness in this context arises when the completion time of a job ($C_j$) exceeds its due date ($d_j$).

To effectively address this scheduling challenge, we explore various parameter settings to identify an optimal generic set. This involves varying the population size, modifying the mutation rate, and experimenting with different constant values for convergence. The evolutionary algorithm, a cornerstone of our approach, incorporates principles of natural evolution into the process of finding optimal solutions for Solver problems. This algorithm directly utilizes the decision variables and problem functions in its methodology. It is noteworthy that evolutionary algorithms are a staple in most commercial Solver applications.

Specifically, in the context of an Excel solver, there are configurable options for evolutionary solving parameters. These options are critical for tailoring the evolutionary algorithm to effectively respond to the unique demands of the scheduling problem at hand, thereby facilitating the discovery of rapid improvements and the best possible solutions.
In this research, specific attention was given to:

**A) Convergence**
The Convergence parameter in Solver specifies the maximum allowable percentage difference in objective values for the system to consider it has "converged to the current solution" for the predominant portion of the population (top 99%). A smaller value in this parameter results in a longer computation time but brings the solution closer to the ideal. In this research, the Convergence was set consistently at two distinct levels: 0.0001, the standard default, and 0.1, to observe the differential impacts on the solution accuracy.

**B) Mutation Rate Parameter**
The Mutation Rate parameter is defined as a value between 0 and 1, representing the frequency at which certain members of the population undergo modification or "mutation" to create new experimental solutions. This rate plays a pivotal role in the evolutionary approach by influencing the examination of each "generation" or subset of the problem. A higher mutation rate increases the population's diversity and the probability of identifying superior solutions, albeit at the cost of extended total solution time. In our study, the mutation rates were set at 0.75, 0.075 (the default), and 0.0075.

### C) Population Size Parameter
The Population Size parameter indicates the number of alternative solution points maintained in the population of candidate solutions at any given time by the Evolutionary approach. In this study, the population sizes selected for evaluation were 100, 50, 25, and 10.

### D) Random Seed Parameter
The Random Seed parameter requires a positive integer, serving as a fixed seed for the random number generator utilized in the evolutionary method. A consistent seed number ensures repeatability of results upon each execution of the Solve command. Leaving this field blank allows for a new seed generation with each execution, leading to variability in the final solution. This study maintains the default setting for this parameter.

### E) Maximum Time without Improvement Parameter
This parameter sets the maximum duration the Evolutionary approach will run without noticeable improvement in the objective value of the best solution in the population. If no improvement is detected within this timeframe, Solver will halt with a message indicating no further improvement is possible. This study does not alter the default setting for this parameter.

### F) Require Bounds on Variables Option
By selecting this option, users inform the Evolutionary approach that all decision variables in the model must have defined lower and upper bounds. Providing bounds improves the performance of the Evolutionary approach, especially when these bounds are as narrow as possible. This study retains the default setting for this option.

### G) Structure of the Study
The remaining sections of this study are organized as follows: Section 3 presents a literature review of the Excel Solver technique in single machine scheduling. Section 4 details the solution methodology. Section 5 further elaborates on this methodology, and Section 6 concludes the study.

## 3. Literature Review

To effectively minimize total tardiness, the single machine total tardiness problem involves scheduling a series of jobs on a single machine. Due to the complexity of this problem, Du, and Leung[1] classified it as NP-hard, signifying that an optimalsolution is not feasible without exhaustive methods. In such cases, heuristic approaches, which aim to find near-optimal solutions efficiently, become valuable alternatives.Solutions to combinatorial problems like this can be categorized as follows:

a) Complete enumeration method.
b) Mathematical modeling.
c) Implicit enumeration/branch-and-bound method.
d) Heuristic method.

While complete enumeration, implicit enumeration, and mathematical modeling methods yield optimal solutions, they are not always practical. For instance, the number of sequences to evaluate in complete enumeration grows exponentially, making it impractical for larger problems. Implicit enumeration methods [2], utilizing branch-and-bound or dynamic programming, are less time-consuming. Innovations in this field include Hirakawa and Yasuhiro's [3] rapid optimal algorithm using branch and bound, and Biskup et al.'s [2] efficient recursion over Kondakci et al.'s [4] branch-and-bound approach. However, these methods can suffer from high temporal complexity.

In contrast, mathematical modeling, though capable of identifying the optimal sequence to minimize total tardiness[5], is limited by the extensive number of variables and constraints, making it suitable only for smaller-scale problems. Heuristic methods, as developed by Wilkerson and Irwin[6], Kim et al.[7], and others, offer more feasible solutions for larger problems. For instance, Tian et al.[8]identified conditions for polynomial solvability in specific cases, and Kiyuzato et al.[9] applied heuristics to real-world scheduling in auto parts manufacturing.

Meta-heuristics, as suggested by Feldmann and Biskup [10] and Cheng et al. [11], further enhance solution quality. These include evolutionary methods, simulated annealing, and Ant Colony Optimization (ACO), offering near-optimal solutions for complex scheduling problems.

Tuning parameters in evolutionary algorithms is crucial for efficient problem-solving. A hybrid parameter tuning approach has been proposed to optimize performance metrics of these algorithms[12]. Bajwa et al. [13] and Cao et al.[14] explored scheduling in group technology systems and production workshops, respectively, using these optimized algorithms.

Burke and Smith discussed the concept of a memetic algorithm, a blend of genetic algorithms and cultural evolution models [15]. Moscato and Norman [2] coined the term 'memetic algorithm' to describe evolutionary algorithms integrated with intensive local search. This approach, based on Dawkins' concept of memes [15], allows for the adaptation and improvement of ideas, different from the rigid transmission of genes.

In this study, an evolutionary algorithm incorporating these principles is employed to address the single machine scheduling problem. This approach directly utilizes decision variables and problem functions in an evolutionary framework [16].

## 4. Methodology

This section of the study details the experimental methodology employed, utilizing the Microsoft Excel Solver's Evolutionary algorithm feature. The computational experiments were conducted on a Lenovo laptop equipped with a dual 2.2 GHz CPU and 16.0 GB of RAM.

Solver, an add-in tool for Microsoft Excel, enables what-if analysis by identifying an optimal value (maximum or minimum) for a formula in a designated cell, referred to as the objective cell. This process is subject to constraints imposed by other formula cells on the worksheet. Solver operates using a group of cells known as choice variables, or variable cells, which are integral in calculating the formulas in both the objective and constraint cells. Through iterative modifications to the values of these choice variable cells, Solver strives to meet the constraints of the constraint cells while achieving the desired result in the objective cell.

In this study, we demonstrate an Excel-based heuristic solution strategy for sequencing problems. For illustrative purposes, we utilize 'Example 1.1' located in sheet 1 of the Excel file. This example, labeled as a T-problem, involves the sequencing of
ten jobs. The experimental setup and the application of the Solver tool in this context aim to provide a practical demonstration of the algorithm's capabilities in addressing complex scheduling challenges.

**Table 1:** T-problem ten jobs

| Job(j) | 1 | 2 | 3 | 4 | 5 | 6 | 7 | 8 | 9 | 10 |
|---|---|---|---|---|---|---|---|---|---|---|
| **Process(pj)** | **11** | **19** | **14** | **10** | **20** | **19** | **19** | **16** | **11** | **14** |
| Due date (dj) | 57 | 58 | 85 | 148 | 100 | 135 | 75 | 94 | 73 | 125 |

In this study, we have meticulously developed a series of modules for an Excel Solver implementation, designed to address various facets of our optimization problem. These facets include problem data organization, the generation of random numbers employing a uniform distribution, task sequencing, performance evaluation, and the execution of essential computations. Figure 1, included in the study, illustrates the standard layout of our model. The key components of this model, essential for the optimization process, are described as follows:

**A) Data Analysis Module**
(Random Number Generation using Discrete Distribution): Situated in Cells S12 to J13, this module is responsible for generating random numbers based on a discrete distribution. This feature is crucial for introducing variability and ensuring the robustness of the model under various data scenarios.

**B) Objective Function**
Located in Cell S35, the objective function is the cornerstone of the model. It defines the goal or the target outcome that the optimization process aims to achieve or maximize/minimize.

**C) Problem Data**
This data is organized in Cells S19 to J21. These cells contain the essential information and parameters that define the specifics of the optimization problem being addressed.

**D) Decision Variables**
These are represented by the sequence in the range S29 to J29. Decision variables are pivotal as they are the elements that the Solver manipulates to find the optimal solution as per the defined objective function.

**E) Applicable Constraints**:
These constraints, which will be defined in subsequent sections of the study, set the boundaries within which the Solver operates. They ensure that the solutions are not only optimal but also feasible within the given problem context.

Each of these components plays a vital role in the functioning of the optimization model. Their careful integration within the Excel Solver framework allows for a systematic and efficient approach to solving the optimization problem, with clear delineation for analysis and interpretation of the outcomes.

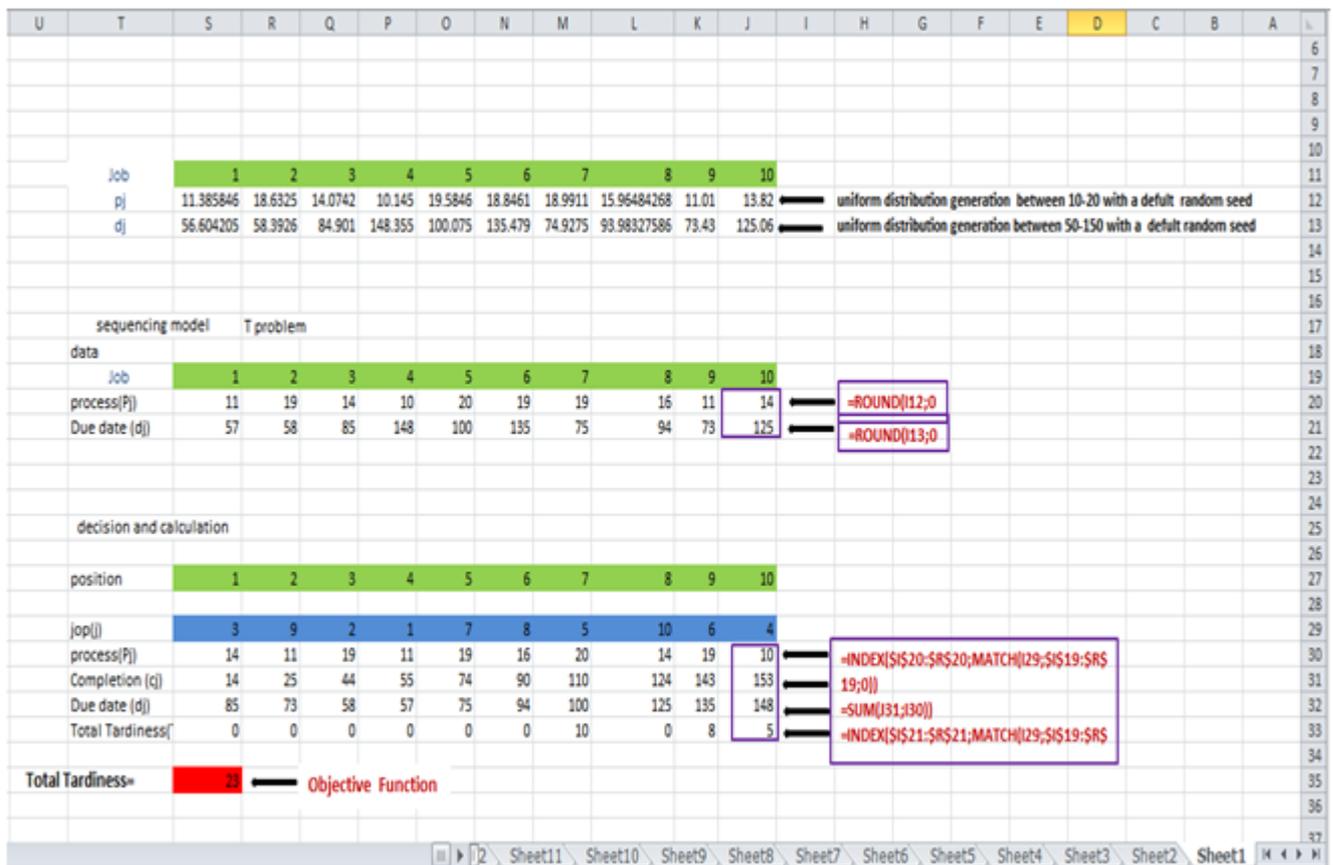

**Figure 1:** Solver Excel model for the T-problem example

In our research, we employed Data Analysis in Excel Solver to generate random numbers based on a discrete distribution. This method was utilized for determining processing times, which ranged between 50-150 units, and due date times, set between 10-20 units, across all problems addressed in the study. Specifically, in Example 1.1, as illustrated in figure 2, we generated random numbers considering the number of variables (jobs) as 10. The generation involved a single row of random numbers, applying a uniform distribution for processing times set between 10-20 units and due date times ranging from 50-150 units. This procedure was executed with a default random seed to maintain consistency and replicability in the results.

For other problems beyond Example 1.1, detailed explanations and methodologies are provided in the corresponding Excel sheets. These sheets serve as a comprehensive repository of data and methodologies applied across different problem scenarios within the scope of this study. The utilization of Excel Solver for random number generation ensures a systematic and standardized approach to creating varied problem sets, thereby enhancing the robustness and applicability of the research findings.

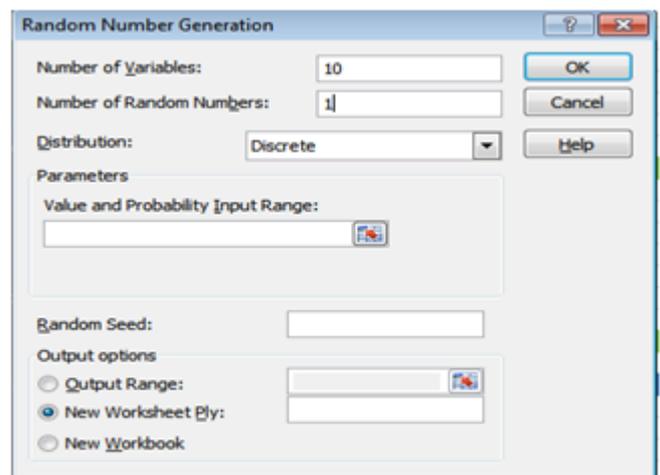

**Figure 2:** Random Number Generation based Solver Excel

In this model, the highlighted cells in row 29 represent decision variables, where permutations of integers 1–10 are inserted to determine job sequences. To ascertain processing times, located in row 30, a lookup algorithm is employed, drawing on the sequence and referencing corresponding data cells. The formula used in cell S30, **INDEX($J$20:$S$20; MATCH (S29; $J$19: $S$19;0))**, exemplifies this approach and is replicated across adjacent cells to the right.

Subsequently, the completion times in row 31 are computed by aggregating the current processing time with the completion time of the preceding job within the sequence. This mirrors the manual calculation method. The due dates in row 32 are similarly deduced, referencing data cells with the formula in cell S32 being **INDEX ($J$21: $S$21;**

**MATCH (S29; $J$19: $S$19;0)**, which is also extended to the right. Each job's lateness, calculated in row 33, utilizes the formula MAX(0;S31-S32) and follows the same right-hand replication.

The objective function, situated in cell S35, is to minimize the sum of tardiness values across all jobs, calculated using the formula SUM(J33:S33). To address the optimization challenge in this example, the focus is on selecting an appropriate sequence in row 29. The tool employed for this purpose is the Excel Solver program, particularly its Evolutionary Solver algorithm. This sophisticated algorithm, one of three available in Solver, is particularly adept at solving complex sequencing problems. To initiate the algorithm, users access the Solver through the add-ins tab, leading to the Solver Parameters box, as demonstrated in Figure 3.

This methodological approach illustrates the application of the Evolutionary Solver in addressing sequencing issues, showcasing its potential in solving complex problems through a systematic and structured procedure.)

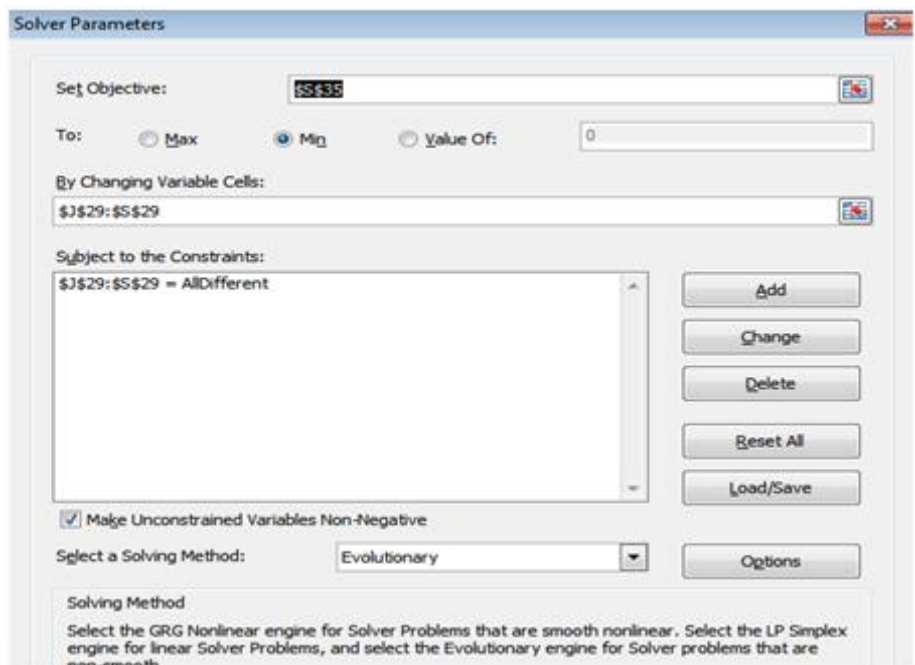

**Figure 3:** Initial Solver windows

In our research, the utilization of the Solver tool in Excel is methodically structured to acquire necessary data by specifically designating parameters. The process involves the following steps:
1) **Setting the Objective Cell**: The cell designated for the objective function, in this case, is Cell B8. This cell is configured to determine the minimum value of the target metric, aligning with the optimization goal of the study.
2) **Defining Variable Cells**: The range of variable cells, identified as Cells C11 to G11, is established to allow Solver to modify these values within specified constraints. This range represents the decision variables crucial to the optimization process.
3) **Applying Constraints**: Constraints are applied to ensure that the solutions provided by Solver remain within feasible and logical bounds. This step involves selecting the objective, variables, and constraints, and then incrementally building the Solver model by clicking the 'Add' button for each element.
4) **Building the Solver Model**: The 'Add Objective' window is accessed by selecting 'Objective' and clicking 'Add'. As depicted in Figure 3, Cell S35 is entered as the objective function cell, with the option set to 'Minimum'. This selection directs Solver to minimize the value in Cell S35. Once 'OK' is clicked, the Solver Parameters window updates to reflect these settings.
5) **Configuring Decision Variables**: As illustrated in Figure 4, the 'Add Variable Cells' window is used to specify the range of decision variables. By setting this range, the Solver is provided with a clear boundary within which it can alter values to find an optimal solution.

Each of these steps plays a critical role in configuring Solver to efficiently address the optimization problem at hand. By methodically setting objectives, defining variable cells, and applying constraints, the Solver model is tailored to seek the most effective solution within the parameters of the study. This systematic approach ensures the accuracy and relevance of the results generated by Solver.

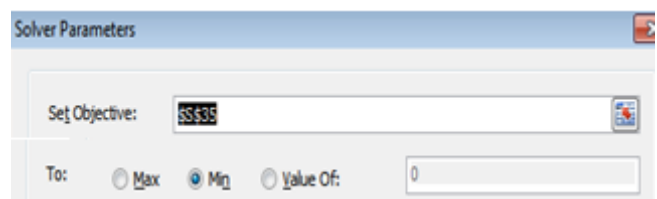

**Figure 4:** Specifying the variables in the example model

as seen in Figure 5, we go to the Add Constraint window. The all-differentconstraints are selected using the pull-down window in the center, and the cell reference corresponds to the range of decision variables. This constraint ensures that the decision variable cells include a valid permutation (in this example, 1–10). To put it another way, the decision cells must follow a logical order. Then, by clicking OK, we return to the Solver Parameters window.

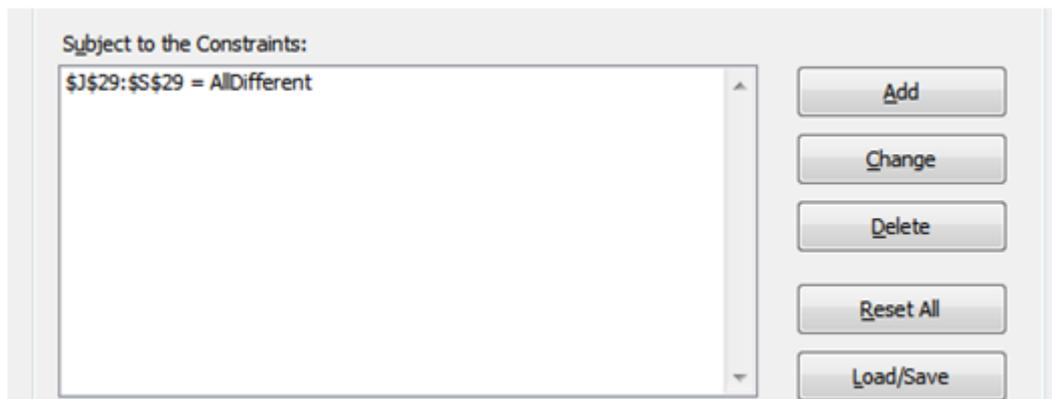

**Figure 5:** Imposing a constraint in the example model

The Solver Parameters box has been adjusted to reflect the problem statement, but one more step remains to complete. As illustrated in Figure 6, we select the Evolutionary Solver as the solution algorithm using pull-down menu.

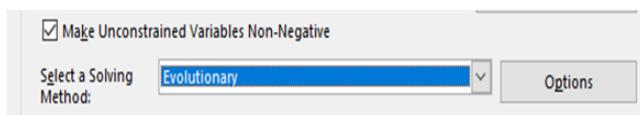

**Figure 6:** Choosing the Evolutionary Solver

In The Evolutionary Solver, It will seek out the optimal solution it can get, and its success is controlled by a number of user-defined parameters that can be set after selecting the Options button in the Solver Parameters box, as shown in figure 7. The most crucial of these variables is

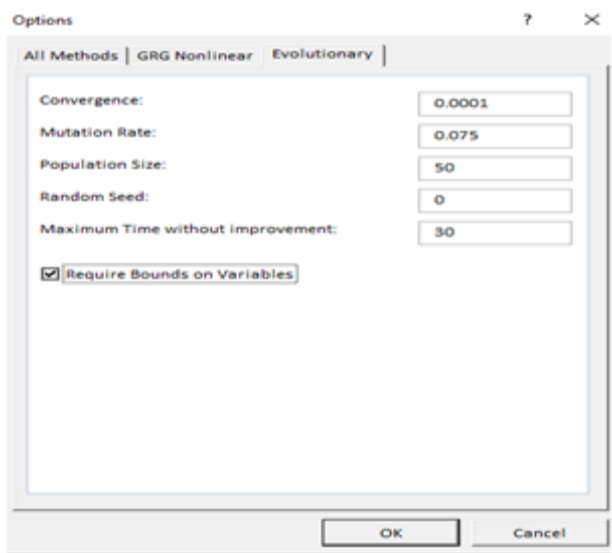

**Figure 7:** Options button in the Solver Parameters window

The convergence conditions and stopping that govern the search's conclusion to achieve a suitable generic collection of parameters, we select some values that are larger and lower than default Values in this search:

Population size is changed = 100, 50, 25, 10.
Mutation rate is changed = 0.75, 0.075 (the default value), 0.0075.
Convergence is constant = 0.0001 (the default value), 0.1.

## 5. Results

Firstly, when we use:
Population size = 100, 50.
Mutation rate = 0.75, 0.075 (the default value), 0.0075.
Convergence = 0.0001 (the default value)

Using parameters 0.75 mutation rate with 100,50 population size ,the search to get the best solution has been taken longer time than other . It took an average about 42-44 seconds, 3-12 seconds at 0.075 mutation rate with 100 population size, but at the shortest time was between 1-4 seconds when using 0.075,0.0075 mutation rate with population size 50 . the improvement and best solutions are found within 0.01% as figure 8

| convergence | population size | mutation Rate | run time to solve problem(seconds) | | | | | | | | | |
|---|---|---|---|---|---|---|---|---|---|---|---|---|
| | | | problem 1 | problem 2 | problem 3 | problem 4 | problem 5 | problem 6 | problem 7 | problem 8 | problem 9 | problem10 |
| 0.0001 | 100 | 0.75 | 42.68 | 42.2 | 43.53 | 41.62 | 43.22 | 41.66 | 41.61 | 43.6 | 43.31 | 41.59 |
| | | 0.075 | 7.58 | 10.44 | 5.21 | 5.38 | 4.98 | 4.02 | 5.93 | 4.66 | 3.73 | 5.67 |
| | | 0.0075 | 5.35 | 5.2 | 4.07 | 3.25 | 3.45 | 2.42 | 2.51 | 3.28 | 2.95 | 2.72 |
| | 50 | 0.75 | 42.1 | 42.4 | 41.99 | 41.13 | 42.14 | 41.83 | 41.48 | 43.49 | 42.21 | 42.76 |
| | | 0.075 | 2.43 | 1.84 | 2.17 | 1.81 | 1.86 | 1.31 | 1.32 | 1.41 | 2.64 | 1.46 |
| | | 0.0075 | 1.97 | 1.47 | 4.07 | 3.25 | 3.45 | 2.42 | 2.51 | 3.28 | 2.95 | 2.72 |
| Solution | | | 23 | 0 | 25 | 8 | 24 | 0 | 1 | 9 | 90 | 7 |
| convergence | population size | mutation Rate | run time to solve problem | | | | | | | | | |
| | | | problem 11 | problem 12 | problem 13 | problem 14 | problem 15 | problem 16 | problem 17 | problem 18 | problem 19 | problem 20 |
| 0.0001 | 100 | 0.75 | 42.76 | 42.15 | 43.29 | 43.08 | 43.79 | 42.45 | 43.89 | 43.58 | 43.93 | 43.95 |
| | | 0.075 | 2.74 | 2.53 | 2.37 | 4.23 | 2.82 | 2.63 | 2.47 | 4.33 | 2.92 | 2.96 |
| | | 0.0075 | 3.26 | 2.29 | 2.46 | 2.83 | 2.92 | 2.49 | 2.66 | 2.93 | 2.99 | 3.19 |
| | 50 | 0.75 | 41.46 | 41.17 | 42.4 | 42.57 | 42.06 | 41.27 | 42.14 | 42.67 | 42.16 | 42.26 |
| | | 0.075 | 1.83 | 1.01 | 1.62 | 2.04 | 1.69 | 1.11 | 1.72 | 2.14 | 1.79 | 1.89 |
| | | 0.0075 | 1.533 | 0.98 | 1.74 | 2 | 2.01 | 1.09 | 1.84 | 2.12 | 2.13 | 2.23 |
| Solution | | | 2 | 2 | 6 | 16 | 19 | 5 | 13 | 33 | 39 | 41 |

**Figure 8:** Population size is (100,50), Mutation rate is 0.075, 0. 75, 0.0075), and Convergence is 0.0001

Secondly, when we use:
Population size = 25, 10.
Mutation rate = 0.75, 0.075 (the default value), 0.0075.
Convergence = 0.0001 (the default value)

The search to get the best solution has been taken average time about 3-13 seconds using parameters 0.75 mutation rate with 25 population size, 1-3 seconds at 0.075 mutation rate, but using parameters (0.075,0.0075) mutation rate with (25,10 ) with constant Convergence at 0.0001 (the default value), the improvement and best solutions cannot find within 0.0001 as figure9

*NA is represented as not found optimal solution.

| convergence | population size | mutation Rate | run time to solve problem(seconds) | | | | | | | | | |
|---|---|---|---|---|---|---|---|---|---|---|---|---|
| | | | problem 1 | problem 2 | problem 3 | problem 4 | problem 5 | problem 6 | problem 7 | problem 8 | problem 9 | problem10 |
| 0.0001 | 25 | 0.75 | 12.79 | 11.59 | 14.76 | 11.18 | 11.93 | 1.15 | 3.25 | 4.88 | 8.35 | 3.46 |
| | | 0.075 | 2.05 | 1.75 | 2.07 | 1.77 | 1.55 | 1.01 | 1.25 | 1.79 | 3.25 | 1.13 |
| | | 0.0075 | NA | 1.54 | 1.69 | NA | 1.65 | 0.98 | 1.09 | 1.51 | NA | 1.17 |
| | 10 | 0.75 | 2.48 | NA | NA | 2.67 | NA | NA | 1.01 | 1.8 | 1.63 | 1.55 |
| | | 0.075 | NA | 1.19 | 3.28 | 1.3 | 1.56 | 0.96 | 1.04 | 1.16 | 1.72 | 2.75 |
| | | 0.0075 | NA | 4.45 | NA | NA | NA | 7.18 | NA | NA | NA | NA |
| Solution | | | 23 | 0 | 25 | 8 | 24 | 0 | 1 | 9 | 90 | 7 |

| | population size | mutation Rate | run time to solve problem | | | | | | | | | |
|---|---|---|---|---|---|---|---|---|---|---|---|---|
| | | | problem 11 | problem 12 | problem 13 | problem 14 | problem 15 | problem 16 | problem 17 | problem 18 | problem 19 | problem 20 |
| 0.0001 | 25 | 0.75 | 8.45 | 1.38 | 4.663 | 3.55 | 2.56 | 1.48 | 4.73 | 3.66 | 2.7 | 2.27 |
| | | 0.075 | 1.27 | 0.99 | 1.22 | 1.24 | 1.64 | 1.15 | 1.32 | 1.34 | 1.79 | 1.9 |
| | | 0.0075 | 1.06 | 0.91 | NA | NA | NA | NA | NA | NA | NA | NA |
| | 10 | 0.75 | 1.28 | 1.03 | NA | NA | NA | NA | NA | NA | NA | NA |
| | | 0.075 | 1.15 | .92 | NA | NA | NA | NA | NA | NA | NA | NA |
| | | 0.0075 | NA | NA | NA | NA | NA | NA | NA | NA | NA | NA |
| Solution | | | 2 | 2 | 6 | 16 | 19 | 5 | 13 | 33 | 39 | 41 |

**Figure 9:** Population size is (25,10), 0. 75, 0.075 the default value, 0.0075, and Convergence is 0.0001 the default value.

The other Experiments Population size is changed = 50, 25, 10.
Mutation rate is changed = 0.75, 0.075 (the default value), 0.0075.
Convergence is constant = 0.1.

Using parameters 0.75 mutation rate with 50 population size , the search to get the best solution has been taken longer time than other as the same when we use 0.0001 convergence previously . It took an average time about 42-44 seconds, 1.5-2 seconds at 0.075,0.0075 mutation rate.

The improvement and best solutions are found within 0.1.But using parameters (0.075,0.0075) mutation rate with (25,10 ) with constant Convergence at 0.1 rate, the improvement and best solutions cannot find within 0.1 as figure10

| convergence | population size | mutation Rate | run time to solve problem(seconds) | | | | | | | | | |
|---|---|---|---|---|---|---|---|---|---|---|---|---|
| | | | problem 1 | problem 2 | problem 3 | problem 4 | problem 5 | problem 6 | problem 7 | problem 8 | problem 9 | problem10 |
| 0.1 | 50 | 0.75 | 42.82 | 41.9 | 42.04 | 40.733 | 41.62 | 4.56 | 43.06 | 42.21 | 41.28 | 41.55 |
| | | 0.075 | 1.56 | 1.7 | 1.83 | 1.81 | 1.72 | 1.22 | 1.41 | 1.58 | 1.29 | 1.66 |
| | | 0.0075 | 1.52 | 1.15 | 1.5 | 1.37 | 1.46 | 1.25 | 1.5 | 1.56 | 1.04 | 1.18 |
| | 25 | 0.75 | 3.27 | 3.26 | 10.88 | 5.5 | 25.93 | 0.85 | 1.58 | 6.42 | 2.25 | 1.22 |
| | | 0.075 | NA | 0.87 | NA | NA | NA | 0.76 | NA | NA | NA | NA |
| | | 0.0075 | NA | NA | NA | NA | NA | 0.772 | NA | NA | NA | NA |
| | 10 | 0.75 | NA | NA | NA | NA | NA | NA | NA | NA | NA | NA |
| | | 0.075 | NA | NA | NA | NA | NA | NA | NA | NA | NA | NA |
| | | 0.0075 | NA | NA | NA | NA | NA | NA | NA | NA | NA | NA |
| | | Solution | 23 | 0 | 25 | 8 | 24 | 0 | 1 | 9 | 90 | 7 |
| | | | problem 11 | problem 12 | problem 13 | problem 14 | problem 15 | problem 16 | problem 17 | problem 18 | problem 19 | problem20 |
| 0.1 | 50 | 0.75 | 41.46 | 41.17 | 42.4 | 42.57 | 42.06 | 41.27 | 42.14 | 42.67 | 42.16 | 42.26 |
| | | 0.075 | 1.83 | 1.01 | 1.62 | 2.04 | 1.69 | 1.11 | 1.72 | 2.14 | 1.79 | 1.89 |
| | | 0.0075 | 1.533 | 0.98 | 1.74 | 2 | 2.01 | 1.09 | 1.84 | 2.12 | 2.13 | 2.23 |
| | 25 | 0.75 | 8.45 | 1.38 | 4.663 | 3.55 | 2.56 | 1.48 | 4.73 | 3.66 | 2.7 | 2.27 |
| | | 0.075 | 1.27 | 0.99 | 1.22 | 1.24 | 1.64 | 1.15 | 1.32 | 1.34 | 1.79 | 1.9 |
| | | 0.0075 | 1.06 | 0.91 | NA | NA | NA | NA | NA | NA | NA | NA |
| | 10 | 0.75 | 1.28 | 1.03 | NA | NA | NA | NA | NA | NA | NA | NA |
| | | 0.075 | 1.15 | .92 | NA | NA | NA | NA | NA | NA | NA | NA |
| | | 0.0075 | NA | NA | NA | NA | NA | NA | NA | NA | NA | NA |
| | | Solution | 2 | 2 | 6 | 16 | 19 | 5 | 13 | 33 | 39 | 41 |

**Figure 10:** Population size is (50,25,10), Mutation rate is 0. 75, 0.075 the default value, 0.0075, and Convergence is 0. 01.

## 6. Conclusion

In this study, we explored the optimization of parameters within the Evolutionary Solver for single machine scheduling problems, aiming to minimize total tardiness. We discovered that time constraints within the solver can be adjusted according to the user's preference. Notably, runs lasting 30 seconds or longer typically yielded satisfactory results for sequencing problems involving up to ten jobs, although optimal or near-optimal solutions were often achieved in considerably less time. The study presents these findings, emphasizing the promising nature of the results in terms of solution quality and speed, particularly when employing an evolutionary algorithm (EA). As detailed in Appendix B, EA is demonstrated proficiency in finding optimal solutions for smaller problems and showed potential for larger problems due to lower computational demands, despite not always yielding the best solutions.

Our experiments with the Evolutionary Solver revealed its capability to rapidly converge on different total tardiness values for various problems. This was observed in a set of 20 problems involving 10 jobs each. When the Solver was run with a time limit of up to 42 seconds, as show in appendix figures 11 and 12, optimal solutions were achieved in all 20 problems, using parameters such as a Population size of 100, 50, 25, a Mutation rate of 0.75, and a Convergence of 0.0001 and 0.01. Interestingly, when the runtime was reduced to between 1-12 seconds, with a Population size of 50, 25, a Mutation rate of 0.075, 0.0075, and the same Convergence rates, the Solver still produced optimal solutions in all cases. These findings, detailed in Appendix A and highlighted in green, indicate that a Population size of 50, Mutation rate of 0.075%, and a Convergence rate of 0.0001 are the optimal parameters for achieving the best solutions within a shorter runtime of 1-3 seconds.

However, when the Population size was reduced further to 25, 10, and the Mutation rate and Convergence remained the same, the Evolutionary Solver did not produce optimal solutions. Attempts to alter the Convergence parameter to 0.000001 and 0.001 yielded results similar to those with 0.0001.

In future research, we aim to modify the objective function to focus on minimizing total weighted tardiness in parallel machine scheduling scenarios with non-zero ready times. This adjustment will be tested with the previously mentioned parameters to ascertain the feasibility of achieving faster optimal solutions.

## 7. Appendix

| Problem | Sequence before improvement (Initial) | | | | | | | | | | solution | Sequence (After improvement to get the best solution) | | | | | | | | | | solution |
|---|---|---|---|---|---|---|---|---|---|---|---|---|---|---|---|---|---|---|---|---|---|---|
| 1  | 3  | 9  | 2 | 1 | 7 | 8  | 5 | 10 | 6 | 4  | 23  | 3  | 9  | 2  | 1 | 7 | 8 | 5  | 10 | 6 | 4 | 23 |
| 2  | 1  | 3  | 2 | 4 | 5 | 7  | 6 | 10 | 9 | 8  | 95  | 10 | 4  | 2  | 6 | 5 | 1 | 8  | 3  | 9 | 7 | 0  |
| 3  | 4  | 2  | 1 | 5 | 7 | 8  | 9 | 3  | 6 | 10 | 212 | 4  | 10 | 6  | 5 | 3 | 2 | 1  | 9  | 7 | 8 | 25 |
| 4  | 4  | 2  | 1 | 5 | 7 | 8  | 9 | 3  | 6 | 10 | 201 | 6  | 10 | 4  | 2 | 3 | 5 | 1  | 9  | 7 | 8 | 8  |
| 5  | 7  | 10 | 8 | 6 | 3 | 2  | 5 | 1  | 4 | 9  | 34  | 7  | 10 | 5  | 6 | 8 | 2 | 3  | 1  | 9 | 4 | 24 |
| 6  | 5  | 4  | 3 | 2 | 1 | 10 | 9 | 8  | 7 | 6  | 60  | 9  | 7  | 10 | 1 | 8 | 6 | 2  | 5  | 3 | 4 | 0  |
| 7  | 10 | 9  | 8 | 7 | 6 | 5  | 4 | 3  | 2 | 1  | 26  | 10 | 2  | 9  | 8 | 6 | 4 | 7  | 1  | 3 | 5 | 1  |
| 8  | 10 | 9  | 8 | 7 | 6 | 5  | 4 | 3  | 2 | 1  | 127 | 1  | 10 | 5  | 7 | 9 | 6 | 8  | 2  | 4 | 3 | 9  |
| 9  | 10 | 9  | 8 | 7 | 6 | 5  | 4 | 3  | 2 | 1  | 202 | 3  | 9  | 10 | 1 | 8 | 5 | 7  | 6  | 4 | 2 | 90 |
| 10 | 10 | 9  | 8 | 7 | 6 | 5  | 4 | 3  | 2 | 1  | 124 | 10 | 3  | 9  | 2 | 1 | 8 | 5  | 4  | 7 | 6 | 7  |
| 11 | 10 | 9  | 8 | 7 | 6 | 5  | 4 | 3  | 2 | 1  | 64  | 8  | 10 | 6  | 3 | 7 | 1 | 9  | 5  | 4 | 2 | 2  |
| 12 | 10 | 9  | 8 | 7 | 6 | 5  | 4 | 3  | 2 | 1  | 85  | 7  | 6  | 10 | 8 | 3 | 4 | 1  | 9  | 2 | 5 | 2  |
| 13 | 10 | 9  | 8 | 7 | 6 | 5  | 4 | 3  | 2 | 1  | 106 | 4  | 9  | 10 | 5 | 6 | 3 | 1  | 2  | 7 | 8 | 6  |
| 14 | 10 | 9  | 8 | 7 | 6 | 5  | 4 | 3  | 2 | 1  | 210 | 10 | 4  | 3  | 5 | 2 | 7 | 8  | 1  | 9 | 6 | 16 |
| 15 | 10 | 9  | 8 | 7 | 6 | 5  | 4 | 3  | 2 | 1  | 105 | 10 | 4  | 6  | 9 | 8 | 7 | 1  | 5  | 2 | 3 | 19 |
| 16 | 10 | 9  | 8 | 7 | 6 | 5  | 4 | 3  | 2 | 1  | 166 | 1  | 7  | 10 | 9 | 4 | 2 | 6  | 8  | 3 | 5 | 5  |
| 17 | 10 | 9  | 8 | 7 | 6 | 5  | 4 | 3  | 2 | 1  | 155 | 1  | 10 | 5  | 4 | 8 | 7 | 3  | 9  | 6 | 2 | 13 |
| 18 | 10 | 9  | 8 | 7 | 6 | 5  | 4 | 3  | 2 | 1  | 120 | 5  | 10 | 9  | 6 | 2 | 4 | 7  | 3  | 8 | 1 | 33 |
| 19 | 10 | 9  | 8 | 7 | 6 | 5  | 4 | 3  | 2 | 1  | 130 | 6  | 10 | 9  | 5 | 4 | 2 | 7  | 3  | 1 | 8 | 39 |
| 20 | 10 | 9  | 8 | 7 | 6 | 5  | 4 | 3  | 2 | 1  | 64  | 5  | 10 | 7  | 8 | 9 | 6 | 4  | 3  | 1 | 2 | 41 |

**Figure 11.1:** 20 Problems solution (the initial and the Optimal (best)solution)

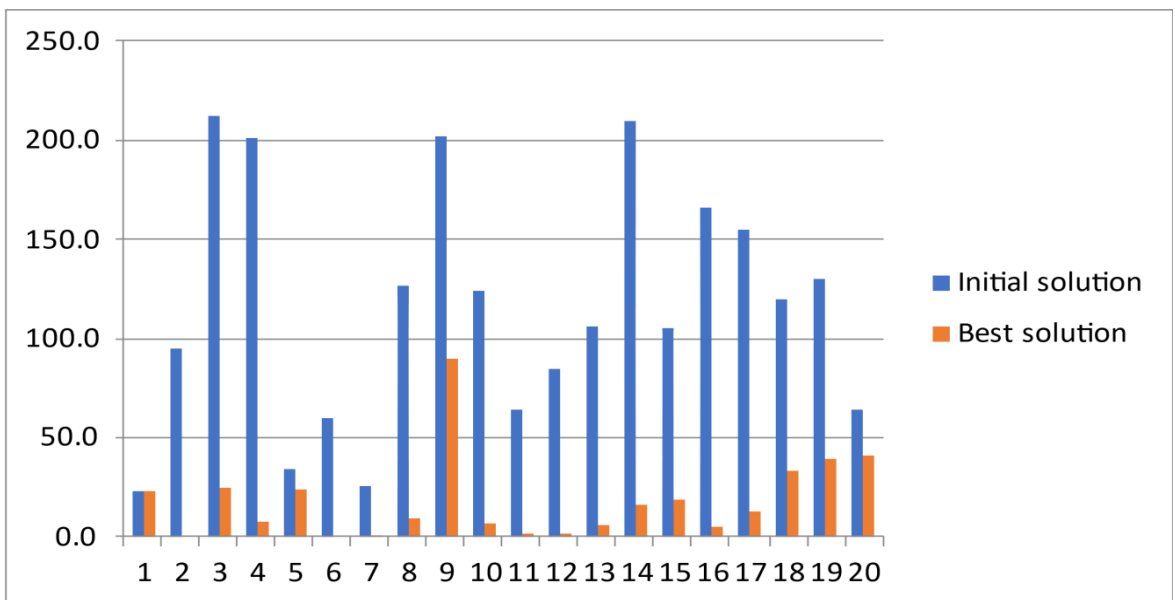

**Figure 11.2:** 20 Problems solution (the initial and the Optimal (best)

| convergence | population size | mutation Rate | run time to solve problem(seconds) | | | | | | | | | |
|---|---|---|---|---|---|---|---|---|---|---|---|---|
| | | | problem 1 | problem 2 | problem 3 | problem 4 | problem 5 | problem 6 | problem 7 | problem 8 | problem 9 | problem10 |
| 0.0001 | 100 | 0.75 | 42.68 | 42.2 | 43.53 | 41.62 | 43.22 | 41.66 | 41.61 | 43.6 | 43.31 | 41.59 |
| | | 0.075 | 7.58 | 10.44 | 5.21 | 5.38 | 4.98 | 4.02 | 5.93 | 4.66 | 3.73 | 5.67 |
| | | 0.0075 | 5.35 | 5.2 | 4.07 | 3.25 | 3.45 | 2.42 | 2.51 | 3.28 | 2.95 | 2.72 |
| | 50 | 0.75 | 42.1 | 42.4 | 41.99 | 41.13 | 42.14 | 41.83 | 41.48 | 43.49 | 42.21 | 42.76 |
| | | 0.075 | 2.43 | 1.84 | 2.17 | 1.81 | 1.86 | 1.31 | 1.32 | 1.41 | 2.64 | 1.46 |
| | | 0.0075 | 1.97 | 1.47 | 4.07 | 3.25 | 3.45 | 2.42 | 2.51 | 3.28 | 2.95 | 2.72 |
| | 25 | 0.75 | 12.79 | 11.59 | 14.76 | 11.18 | 11.93 | 1.15 | 3.25 | 4.88 | 8.35 | 3.46 |
| | | 0.075 | 2.05 | 1.75 | 2.07 | 1.77 | 1.55 | 1.01 | 1.25 | 1.79 | 3.25 | 1.13 |
| | | 0.0075 | NA | 1.54 | 1.69 | NA | 1.65 | 0.98 | 1.09 | 1.51 | NA | 1.17 |
| | 10 | 0.75 | 2.48 | NA | NA | 2.67 | NA | NA | 1.01 | 1.8 | 1.63 | 1.55 |
| | | 0.075 | NA | 1.19 | 3.28 | 1.3 | 1.56 | 0.96 | 1.04 | 1.16 | 1.72 | 2.75 |
| | | 0.0075 | NA | 4.45 | NA | NA | NA | 7.18 | NA | NA | NA | NA |
| 0.1 | 50 | 0.75 | 42.82 | 41.9 | 42.04 | 40.733 | 41.62 | 4.56 | 43.06 | 42.21 | 41.28 | 41.55 |
| | | 0.075 | 1.56 | 1.7 | 1.83 | 1.81 | 1.72 | 1.22 | 1.41 | 1.58 | 1.29 | 1.66 |
| | | 0.0075 | 1.52 | 1.15 | 1.5 | 1.37 | 1.46 | 1.25 | 1.5 | 1.56 | 1.04 | 1.18 |
| | 25 | 0.75 | 3.27 | 3.26 | 10.88 | 5.5 | 25.93 | 0.85 | 1.58 | 6.42 | 2.25 | 1.22 |
| | | 0.075 | NA | 0.87 | NA | NA | NA | 0.76 | NA | NA | NA | NA |
| | | 0.0075 | NA | NA | NA | NA | NA | 0.772 | NA | NA | NA | NA |
| | 10 | 0.75 | NA | NA | NA | NA | NA | NA | NA | NA | NA | NA |
| | | 0.075 | NA | NA | NA | NA | NA | NA | NA | NA | NA | NA |
| | | 0.0075 | NA | NA | NA | NA | NA | NA | NA | NA | NA | NA |
| Solution | | | 23 | 0 | 25 | 8 | 24 | 0 | 1 | 9 | 90 | 7 |

| convergence | population size | mutation Rate | problem |||||||||| 
|---|---|---|---|---|---|---|---|---|---|---|---|
| | | | run time to solve problem |||||||||
| | | | problem 11 | problem 12 | problem 13 | problem 14 | problem 15 | problem 16 | problem 17 | problem 18 | problem 19 | problem 20 |
| 0.0001 | 100 | 0.75 | 42.76 | 42.15 | 43.29 | 43.08 | 43.79 | 42.45 | 43.89 | 43.58 | 43.93 | 43.95 |
| | | 0.075 | 2.74 | 2.53 | 2.37 | 4.23 | 2.82 | 2.63 | 2.47 | 4.33 | 2.92 | 2.96 |
| | | 0.0075 | 3.26 | 2.29 | 2.46 | 2.83 | 2.92 | 2.49 | 2.66 | 2.93 | 2.99 | 3.19 |
| | 50 | 0.75 | 41.46 | 41.17 | 42.4 | 42.57 | 42.06 | 41.27 | 42.14 | 42.67 | 42.16 | 42.26 |
| | | 0.075 | 1.83 | 1.01 | 1.62 | 2.04 | 1.69 | 1.11 | 1.72 | 2.14 | 1.79 | 1.89 |
| | | 0.0075 | 1.533 | 0.98 | 1.74 | 2 | 2.01 | 1.09 | 1.84 | 2.12 | 2.13 | 2.23 |
| | 25 | 0.75 | 8.45 | 1.38 | 4.663 | 3.55 | 2.56 | 1.48 | 4.73 | 3.66 | 2.7 | 2.27 |
| | | 0.075 | 1.27 | 0.99 | 1.22 | 1.24 | 1.64 | 1.15 | 1.32 | 1.34 | 1.79 | 1.9 |
| | | 0.0075 | 1.06 | 0.91 | NA | NA | NA | NA | NA | NA | NA | NA |
| | 10 | 0.75 | 1.28 | 1.03 | NA | NA | NA | NA | NA | NA | NA | NA |
| | | 0.075 | 1.15 | ..92 | NA | NA | NA | NA | NA | NA | NA | NA |
| | | 0.0075 | NA | NA | NA | NA | NA | NA | NA | NA | NA | NA |
| 0.1 | 50 | 0.75 | 41.84 | 41.2 | 42.28 | 42.78 | 43.93 | 41.12 | 42.38 | 42.88 | 43.98 | 43.99 |
| | | 0.075 | 1.63 | 1.34 | 1.78 | 1.97 | 1.54 | 1.44 | 1.88 | 1.99 | 1.64 | 1.67 |
| | | 0.0075 | 1.04 | 1.31 | 1.68 | 1.66 | 1.36 | 1.41 | 1.78 | 1.76 | 1.46 | 1.56 |
| | 25 | 0.75 | 5.94 | 4.94 | 13.82 | 2.83 | 9.53 | 4.97 | 13.92 | 2.93 | 9.63 | 6.69 |
| | | 0.075 | NA | NA | NA | NA | NA | NA | NA | NA | NA | NA |
| | | 0.0075 | NA | NA | NA | NA | NA | NA | NA | NA | NA | NA |
| | 10 | 0.75 | NA | NA | NA | NA | NA | NA | NA | NA | NA | NA |
| | | 0.075 | NA | NA | NA | NA | NA | NA | NA | NA | NA | NA |
| | | 0.0075 | NA | NA | NA | NA | NA | NA | NA | NA | NA | NA |
| Solution | | | 2 | 2 | 6 | 16 | 19 | 5 | 13 | 33 | 39 | 41 |

**Figure 11.3:** Optimal Solver Excel Parameters for 20 problems

| Problem | Best solution | Initial solution |
|---|---|---|
| 1 | 23.0 | 23.0 |
| 2 | 0.0 | 95.0 |
| 3 | 25.0 | 212.0 |
| 4 | 8.0 | 201.0 |
| 5 | 24.0 | 34.0 |
| 6 | 0.0 | 60.0 |
| 7 | 1.0 | 26.0 |
| 8 | 9.0 | 127.0 |
| 9 | 90.0 | 202.0 |
| 10 | 7.0 | 124.0 |
| 11 | 2.0 | 64.0 |
| 12 | 2.0 | 85.0 |
| 13 | 6.0 | 106.0 |
| 14 | 16.0 | 210.0 |
| 15 | 19.0 | 105.0 |
| 16 | 5.0 | 166.0 |
| 17 | 13.0 | 155.0 |
| 18 | 33.0 | 120.0 |
| 19 | 39.0 | 130.0 |
| 20 | 41.0 | 64.0 |

**Figure 121.**4 Initial and Optimal solution for 20 problems